\documentclass{osa-article}

\journal{osajournal}

\articletype{Research Article}

\usepackage{lineno}
 \usepackage{relsize}

\newcommand{\D}{\mathrm{d}}

\begin{document}

\title{High Dose-Rate Ionizing Radiation Source from Tight Focusing in Air of a mJ-class Femtosecond Laser}

\author{Simon Valli\`eres\authormark{1,2,$\dagger$,*}, Jeffrey Powell\authormark{1,$\dagger$}, Tanner Connell\authormark{3}, Michael Evans\authormark{3}, Sylvain Fourmaux\authormark{1}, St\'ephane Payeur\authormark{1}, Philippe Lassonde\authormark{1}, Fran\c cois Fillion-Gourdeau\authormark{4}, Steve MacLean\authormark{1,2,4} and Fran\c cois L\'egar\'e\authormark{1}}

\address{\authormark{1} INRS-EMT, 1650 blvd. Lionel-Boulet, Varennes, QC, J3X 1P7, Canada\\
\authormark{2} Institute for Quantum Computing, University of Waterloo, Waterloo, ON, N2L 3G1, Canada\\
\authormark{3} Medical Physics Unit, McGill University Health Center, 1001 blvd. D\'ecarie, Montr\'eal, QC, H4A 3J1, Canada\\
\authormark{4} Infinite Potential Laboratories, Waterloo, ON, N2L 0A7, Canada}
\hspace{-3.5mm}\authormark{$\dagger$}\textit{ \small Equivalent first-author contribution}\\
\email{\authormark{*} Corresponding author: simon.vallieres@uwaterloo.ca} 



\begin{abstract}
Ultrashort electron beams with femtosecond to picosecond bunch durations offer unique opportunities to explore active research areas ranging from ultrafast structural dynamics to ultra-high dose-rate radiobiological studies. We present a straightforward method to generate MeV-ranged electron beams in ambient air through the tight focusing of a few-cycle, mJ-class femtosecond IR laser. At one meter from the source, the highest measured dose rate of 0.15 Gy/s exceeds the yearly dose limit in less than one second and warrants the implementation of radiation protection.  Two-dimensional Particle-In-Cell simulations confirm that the acceleration mechanism is based on the relativistic ponderomotive force and show theoretical agreement with the measured electron energy. Furthermore, we discuss the scalability of this method with the continuing development of mJ-class high average power lasers, moreover providing a promising approach for FLASH radiation therapy.
\end{abstract}

\section{Introduction}
\noindent

Since Tajima and Dawson\cite{Tajima1979} theoretically proposed the ponderomotive force to generate strong accelerating fields in plasmas, the field of electron acceleration driven by high intensity lasers has rapidly progressed. The advent of Chirped Pulse Amplification (CPA)\cite{Strickland1985} gave rise to laser wakefield acceleration (LWFA)\cite{Lu2007,Esarey2009} which is now frequently achieved with Ti:Sapphire lasers with peak powers up to a few PW. Current advances of this technique have enabled LWFA to work with mJ-class lasers at kHz repetition rates\cite{Faure2019} and in the mid-IR\cite{Woodbury2018}. Taking advantage of the high particle flux that these high-average power lasers can generate, LWFA has recently been used in an experiment to irradiate biological cells with electron beam energies up to 2 MeV\cite{Cavallone2021}. \\

\noindent The significant increase in the radiation dose from laser-driven ionizing radiation sources requires the implementation of the appropriate radiation protection safety measures to avoid potential exposure. Shielding is an expected precaution for experiments performed with Joule-class lasers and this becomes increasingly important with the high repetition rate of mJ-class lasers. The work of Cavallone \textit{et al.}\cite{Cavallone2021} produced an average dose rate in the Gy/s range with a mJ-class, kHz system whereas Joule-class lasers typically produce Gy/min dose rate levels\cite{Lundh2012,Svendsen2021}. In Canada, the public absorbed dose limit is set to 1 mSv per year\cite{cnsc} (equivalent to 1 mGy per year for electrons and photons), emphasizing the need for adequate radiation protection. Hard X-ray sources produced using copper in ambient air and operated at kHz repetition rates have also been demonstrated\cite{Hou2008,Pikuz2010}, making them very appealing due to the inherent simplicity of the experimental setup. However, laser-based in-air configurations require further radiation safety considerations as this simple implementation can lead to an even higher radiation exposure risk.\\

\noindent In this work, we report on the generation of a high dose-rate ionizing radiation source from the tight focusing in air of a mJ-class femtosecond laser. A directional electron beam with energies up to $\sim$1 MeV and effective X-ray energies of $\sim$20 keV were produced. The simple in-air geometry of the setup demonstrates a significant radiation safety concern in a laboratory environment as non-negligible dose rates of ionizing radiation were measured up to 6 m away from the interaction volume. Three different types of radiation detectors with absolute dose calibrations were used in order to separately confirm the data. Dose measurements were collected over 8 orders of magnitude using parameter scans to characterize the high energy electron and X-ray beams, and to determine the underlying dose scaling laws. Finally, the results are further supported by 2D Particle-In-Cell (PIC) simulations which demonstrate the relativistic ponderomotive force as the underlying acceleration mechanism, due to the ability to sustain high intensity IR pulses in ambient air conditions.\\

\section{Methodology}
\label{metho}

\subsection{The ALLS infrared beamline}
\noindent

\begin{figure}[h!]
\centering\includegraphics[width=310pt]{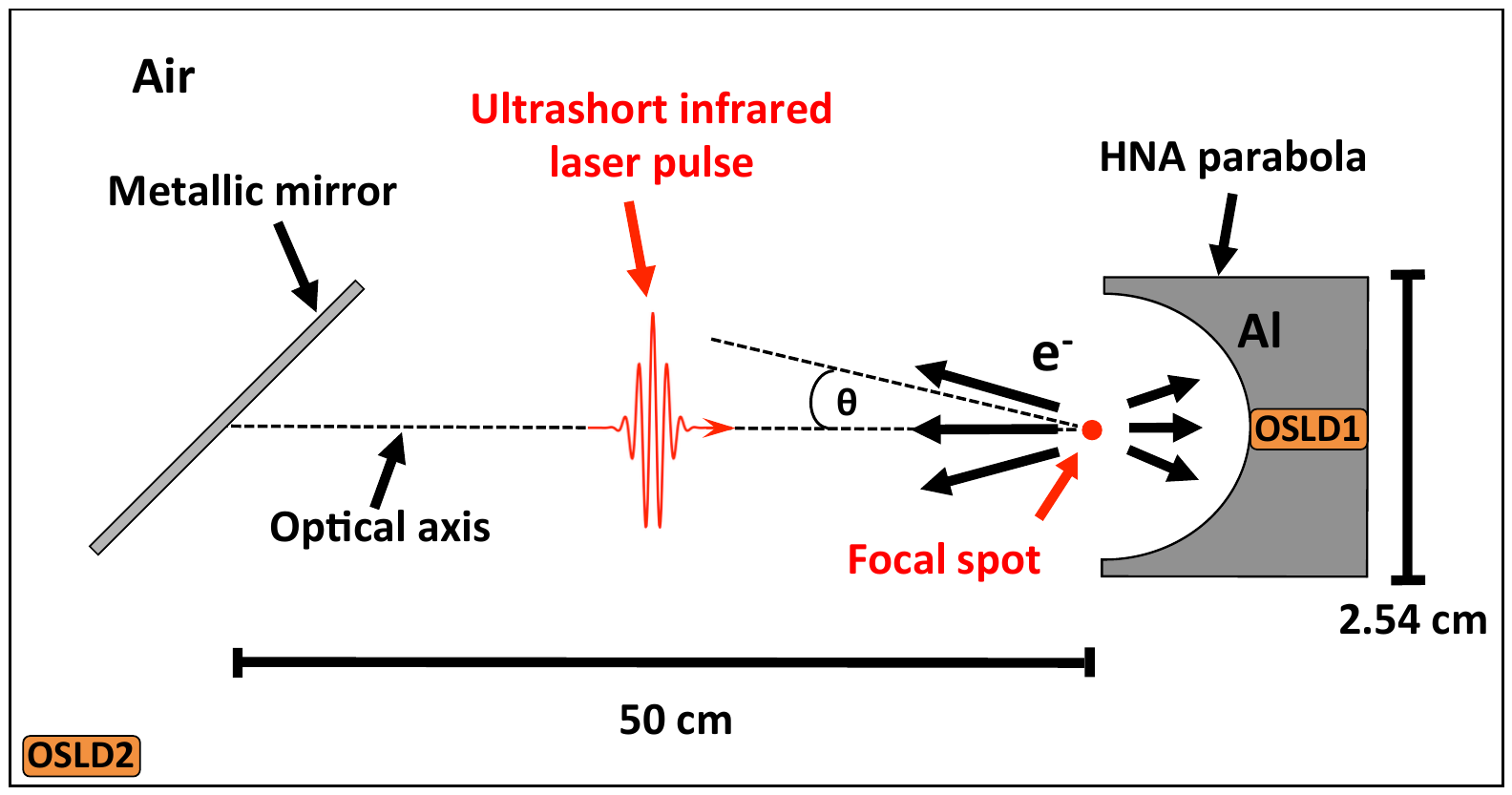}
\caption{Top view sketch of the experimental setup for the in-air, laser-based radiation generation. As the High Numerical Aperture (HNA) focusing parabola is on-axis, the last turning mirror blocks the radiation emitted for conical angles $\theta<4^\circ$ with respect to the optical axis. OSLD1 is placed 0.7 m above (\textit{i.e.}, $\theta=90^\circ$) the focusing optic, whereas OSLD2 is positioned 2 m away from the interaction point at about $\theta=45^\circ$ in the incidence plane of the laser.}
\label{fig1}
\end{figure}

The experiment was performed at the Advanced Laser Light Source (ALLS) facility (Varennes, Canada) which provided the few-cycle, IR beamline based on a high-energy Optical Parametric Amplification (OPA). The same system was used for Longitudinal Electron Acceleration (LEA) with radially polarized beams\cite{Payeur2012,Powell2021}. A schematic of the experimental setup is shown in Figure \ref{fig1}. The high-energy OPA\cite{Thire2015} delivers 7 mJ per pulse at 1.8 $\mu$m which is coupled into a 1 mm-diameter, 4.35 m-long, stretched hollow-core fiber filled with a static pressure of 180 Torr of argon\cite{Cardin2015}. After the spectrally broadened output of the hollow-core fiber, dispersion compensation was achieved using fused silica windows\cite{Schmidt2010}. The compressed pulses were characterized using SHG-FROG (Second Harmonic Generation - Frequency-Resolved Optical Gating) yielding two optical cycles with a 12 fs pulse duration at Full-Width-Half-Maximum (FWHM). A ND-filter wheel was used to vary the energy $\mathcal{E}_\textrm{L}$ of the 10.5 mm diameter (FWHM) s-polarized laser beam on the focusing optic ($\mathcal{E}_\textrm{L}=$ 1.2, 1.7, 2.3, 2.7 and 2.8 mJ). The laser repetition rate was 100 Hz with a pulse-to-pulse energy stability of 2.5\% RMS. The tight focusing optic (Fig.\ref{fig1}.A) was an on-axis parabola with focal length of 6.35 mm and diameter of 2.54 cm giving NA $\approx1$. The focal spot is located in ambient air. A metallic turning mirror oriented at 45$^\circ$ was located 50 cm from the parabola focus which defined the optical axis, as shown in Figure \ref{fig1}.B. Note this mirror obstructed the view for $\theta<4^\circ$ making radiation measurements in this range not possible.\\

\subsection{Dosimetry}
\noindent

Three different types of radiation detectors with absolute dose calibrations were used in order to separately confirm the data. Ionizing radiation dose was measured using parametric scans with one of two detector types depending on the dose rate at the measurement point. For the higher dose rate measurements close to the source, a classical Farmer-type air-filled ionization chamber was used as Detector 1 (D1). The calibration coefficients for this Exradin A12 ion chamber were determined in a primary standards laboratory using a SuperMax electrometer to collect the charge. This total collected charge was converted to the absolute dose using the AAPM TG-51 formalism\cite{tg51} corresponding to precision-level clinical dosimetry. For the purpose of our radiation protection survey, the temperature and pressure corrections were important and taken into account for all measurements. All other correction factors in this formalism were set to unity as the impact on the results would be negligible. For the lower dose measurements farther from the source, a calibrated Fluke 451B hand-held portable survey meter was used as Detector 2 (D2) because D1's active volume of 0.64 cm$^3$ is not sensitive enough. D2 has a very sensitive pressurized volume of 349 cm$^3$ with a thin entrance window that can detect low energy electrons ($\geq 100$ keV) and photons ($\geq 7$ keV). For all measurements, both D1 and D2 were operated in integration mode and no operator was required to be present during the acquisition. Each measurement was made with the detector placed in an un-obstructed view of the focal spot at about $\theta=20^\circ$ with respect to the optical axis. Each measurement was recorded for one minute at distances of $r=$ 0.1, 0.15, 0.25, 0.5, 1, 2, 3, 4, 5 and 6 m between the source and the detector.\\

\noindent The third detector type consisted of two identical Optically-Stimulated Luminescent Devices (OSLD1 and OSLD2) from \textit{Health Canada}\cite{healthcanada}. These were used as area dosimeters to record the accumulated dose over the course of the experiment. OSLD1 was placed 0.7 m above the focusing optic ($\theta=90^\circ$), and OSLD2 was placed 2 m away from the focus at an angle of $\theta=45^\circ$. Each detector reports two dosimetric quantities, namely the "surface dose" H$_\textrm{p}$(0.07) and the "body dose" H$_\textrm{p}$(10) which are equivalent dose levels at 0.07 and 10 mm depths in water, respectively. In addition, this type of dosimeter was always worn by each person attending the experiment.\\

\section{Results} \label{res}
\noindent

Reference dose measurements were taken at a fixed position close to the source (68 cm away at $\theta=45^\circ$) to account for variations such as parabola alignment and laser pulse energy drift, enabling a proper comparison between acquisitions. In addition, a repeatability check was performed on five identical measurements, yielding a standard deviation of 1.1\%. Since detectors D1 and D2 exhibit different energy-dependent sensitivity ranges, we performed a cross-calibration consisting of four identical measurements at the same endpoints (same laser pulse energy and position). This difference in energy sensitivity is explained by the different wall thickness that electrons and photons need to cross to reach the sensitive volume. The ionization chamber (D1), with its 0.5 mm C552 plastic wall has a water-equivalent thickness of 0.88 mm, whereas the survey meter (D2) has a much thinner wall consisting of 0.048 mm of Mylar which equates to a water-equivalent thickness of 0.066 mm. Therefore, it is expected that D1 would underrespond in comparison with D2 as some of the lowest energy electrons would not be able to cross the thicker wall of the ionization chamber. The mean sensitivity scaling factor used to adjust the underresponse of the ionization chamber was considered as the mean value of dose ratios between the two detectors at these four measurement endpoints.

\begin{figure*}[h!]
\centering\includegraphics[width=380pt]{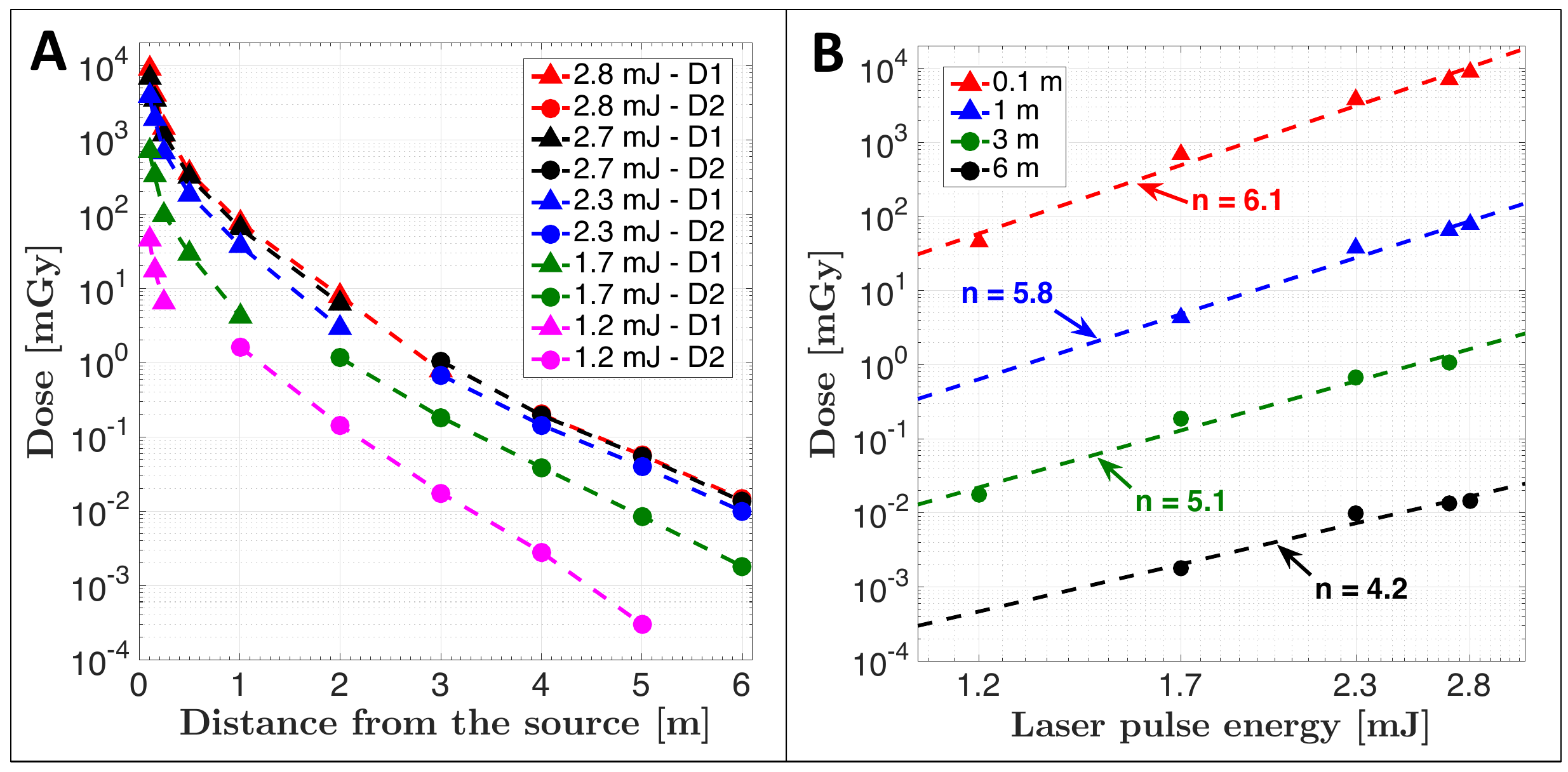}
\caption{Measured radiation doses integrated for 1 min ($\sim20^\circ$ from the optical axis) for the two detectors, D1 (triangles) and D2 (dots) as function of (A) distance from the source, following a typical Inverse-Square Law (ISL), for different laser pulse energies of 1.2 mJ (magenta), 1.7 mJ (green), 2.3 mJ (blue), 2.7 mJ (black) and 2.8 mJ (red). Data points (markers) are connected for better visualization (dashed lines). Radiation doses are also shown as a function of (B) laser pulse energy, for different distances of 0.1 m (red), 1 m (blue), 3 m (green) and 6 m (black). Data points were fitted (dashed lines) with a Power Law of the form $D = a\mathcal{E}_\textrm{L}^\textrm{n}$ using a linear regression in the log-log domain, fitting with $R^2$ = 0.98, 0.98, 0.97 and 0.95 for 0.1 m, 1 m, 3 m and 6 m, respectively.}
\label{fig2}
\end{figure*}

\noindent Figure \ref{fig2}.A displays the measured radiation dose in mGy/min (1 Gy = 1 J/kg) as a function of the distance $r$ from the source, for five different laser pulse energies. It can be seen that the dose follows an Inverse-Square Law (ISL), \textit{i.e.} $D \propto 1/r^2$ fitting with $R^2 >0.99$, as expected for a diverging particle beam emerging from a point-like source. The lowest average dose rates $\dot{\overline{D}}$ reported are on the order of a few $\mu$Gy/min when measured far from the source (5-6 m) at 1.2 mJ, whereas the highest measured dose rate at 0.1 m and 2.8 mJ reaches $\dot{\overline{D}}=$ 8.9 Gy/min (0.15 Gy/s). The data between D1 and D2 was normalized, as discussed above, but nonetheless minor dose steps are noticeable in some of the curves since the cross-calibration points are measuring different particle spectra (type and energy), demonstrating slightly different energy-dependent sensitivity ratios from both detectors.\\

\noindent The radiation dose scaling is shown in Figure \ref{fig2}.B and plotted as a function of laser pulse energy. A power law of the form $D = a\mathcal{E}_\textrm{L}^\textrm{n}$ was fitted in the log-log domain (\textit{i.e.}, linear fits) for all cases. We observed that dose $D$ scales with $\mathcal{E}_\textrm{L}^\textrm{6}$ when measured close to the source (\textit{i.e.}, electrons and X-rays contributing), and slightly decreases to $\mathcal{E}_\textrm{L}^\textrm{4}$ at 6 m (\textit{i.e.}, only X-ray photons contributing). This very strong dose-energy scaling regime indicates that we have not yet reached the saturation, though we expect that non-linear propagation effects and intensity clamping in air will limit the conversion efficiency at higher peak intensities\cite{Becker2001,Liu2002,Kosareva2009,Kandidov2011}. This could limit the scaling of the dose with laser energy and depends on parameters such as the incident wavelength, pulse duration as well as the gas type and pressure.

\begin{figure*}[h!]
\centering\includegraphics[width=380pt]{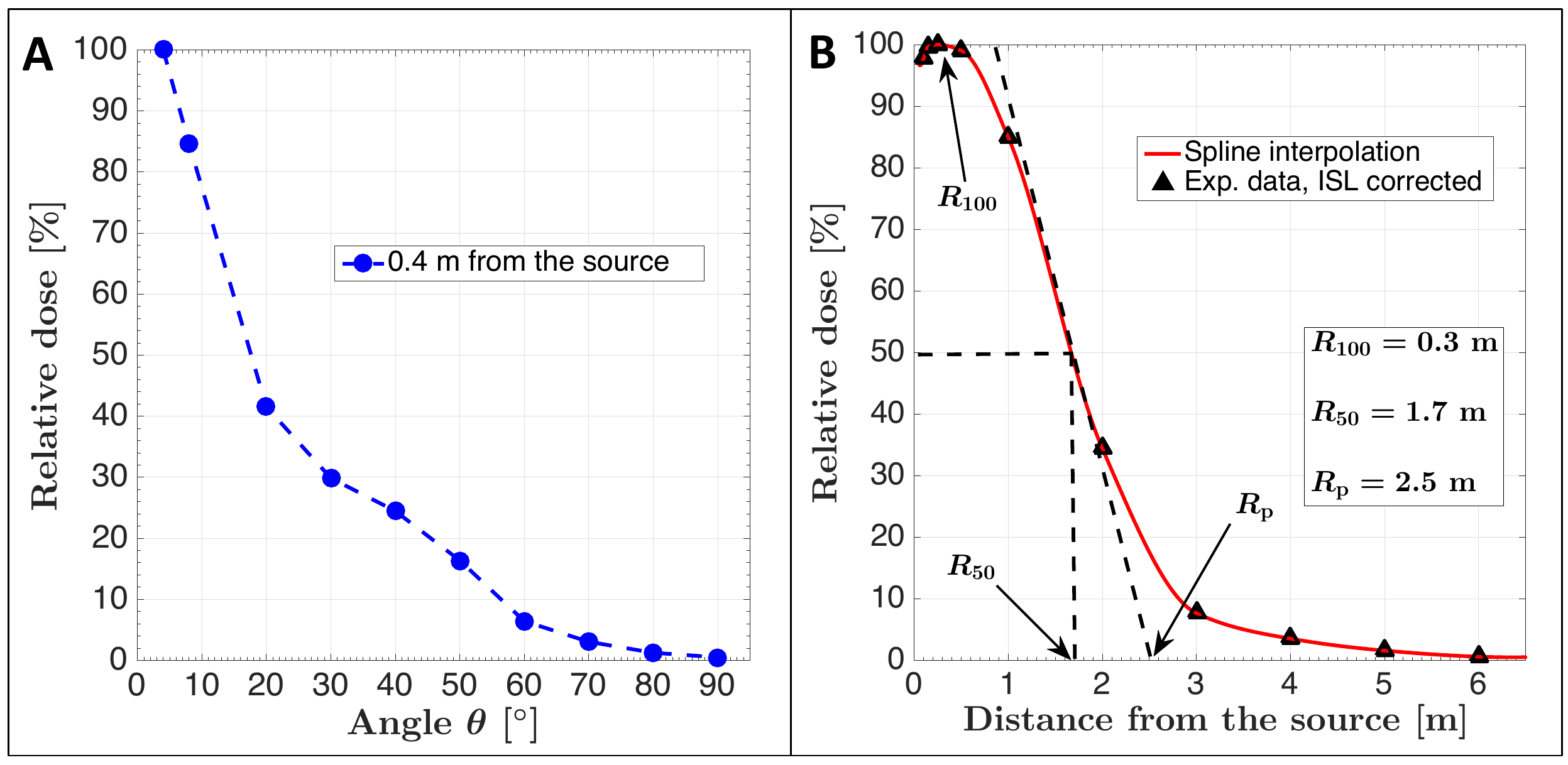}
\caption{Relative dose distributions at highest pulse energy ($\mathcal{E}_\textrm{L}$ = 2.8 mJ) as function of (A) angle $\theta$ with respect to the optical axis as measured 0.4 m from the source (blue dashed line shown for improved visualization of the data), and (B) distance from the source in air, corrected for the Inverse-Square Law. Electron ranges of $R_\textrm{100}$ = 0.3 m and $R_\textrm{50}$ = 1.7 m are obtained using a shape-preserving spline interpolation (full red line). The practical range $R_\textrm{p}$ = 2.5 m is extracted from the tangent line passing through $R_\textrm{50}$, as conventionally done with electron PDD curves.}
\label{fig3}
\end{figure*}

\noindent Further characterization of the radiation source examined the angular dose distribution with respect to the optical axis. Figure \ref{fig3}.A shows the angular dose distribution acquired at 0.4 m from the source and measured for one angular quadrant between 0$^\circ$ and 90$^\circ$. Note the highly peaked distribution (\textit{i.e.}, anisotropic emission) which exhibits an estimated half-cone divergence angle of $\theta_\textrm{HWHM} = 17^\circ$. As the data at $\theta=0^\circ$ was inaccessible due to view obstruction from the last turning mirror, a more realistic estimation would be $\theta_\textrm{HWHM} \leq 17^\circ$. In addition, electrons scattering in air, enroute to the detector, is a source of increased divergence. Figure \ref{fig3}.B plots the relative dose with respect to distance in air for the highest laser energy of 2.8 mJ. The shape of the curve is similar to an electron beam Percent Depth-Dose (PDD) curve that describes the dose deposition profile in a medium as a function of depth. These values were obtained by correcting the dose data points for the ISL in order to extract the dose decrease due to beam divergence (\textit{i.e.}, representing the equivalent of a collimated electron beam), followed by normalizing to the maximum value. This permits a better evaluation of the dose deposition behavior with respect to depth in a material (air in this case), which is characteristic for a particular particle type and its energy. The "electron-like" shape of the curve is characterized by a slight increase at low depths, up to $R_\textrm{100} = 0.3$ m here, and then a rather steep decrease exhibiting a half-dose range of $R_\textrm{50} = 1.7$ m and practical range of $R_\textrm{p} = 2.5$ m. Beyond $r = 3$ m, the dose is characterized by a slowly decreasing tail consisting of Bremsstrahlung X-ray photons, again as expected for an electron PDD consisting of both electrons and X-rays. It is important to note that electron PDDs are typically reported for a monoenergetic electron beam and will not be directly comparable to the broadband electron spectrum obtained in this work. This makes the reported curve slightly different from a typical electron PDD but still gives valuable insight into the reported electron ranges ($R_\textrm{100}$, $R_\textrm{50}$ and $R_\textrm{p}$). The Bremsstrahlung tail on the reported PDD in Figure \ref{fig3}.B stems from the electron irradiation of components on the optical table and radiative losses in air. Nevertheless, since the depth-dose distribution appears to be dominated by the electron population and to a lesser extent by the X-ray photons, it is possible to determine a lower bound on the maximum electron energy of the spectrum using the practical range $R_\textrm{p}$ value and the CSDA (Continuous Slowing Down Approximation) electron range tables in air from the NIST-ESTAR database\cite{nist2}. We estimate the maximum electron energy at 6 m from the source (\textit{i.e.}, the maximum range $R_\textrm{max}$), as observed from the Bremsstrahlung tail in Figure \ref{fig3}.B, thereby delimiting the upper bound of electron energies using CDSA ranges. This leads to an estimated maximum electron energy in the range of $0.8 \leq \mathcal{E}_{\textrm{e}}^\textrm{max} \leq 1.4$ MeV.\\

\noindent In order to verify the validity of the upper bound for the maximum electron energy, we simulated the tightly-focused electromagnetic (EM) fields of the laser using an in-house code that calculates the Stratton-Chu integral formulation of EM fields\cite{Dumont2017}, and takes as input the parameters of the incident laser field and the focusing parabola (see \textit{Supplement 1} for more information). The simulation considers an ideal linearly-polarized beam (\textit{i.e.} no aberrations) and the propagation of the fields in vacuum, therefore not taking into account any possible non-linear effects or plasma generation during the focusing. A spot size of 1.0 $\mu$m at FWHM was obtained, yielding a peak intensity of $I_0\approx 1\times10^{19}$ W/cm$^2$, hence a normalized amplitude of the potential vector $a_0=\sqrt{I_0\lambda_0^2/(1.37\times10^{18} \textrm{ }[\textrm{W}\cdot\mu\textrm{m}^2/\textrm{cm}^2])} = 4.86$. Considering a purely ponderomotive acceleration of the electrons, this gives a cycle-averaged ponderomotive electron energy of $\overline{\mathcal{E}}_{\textrm{e}}^\textrm{pond} = m_\textrm{e} c^2\left( \sqrt{1+a_0^2/2} -1\right) = 1.3$ MeV, in good agreement with the estimated upper bound of 1.4 MeV. The lower bound of 0.8 MeV requires an intensity $I_0 > 4\times10^{18}$ W/cm$^2$ ($a_0 = 3.08$).

\begin{figure*}[h!]
\centering\includegraphics[width=380pt]{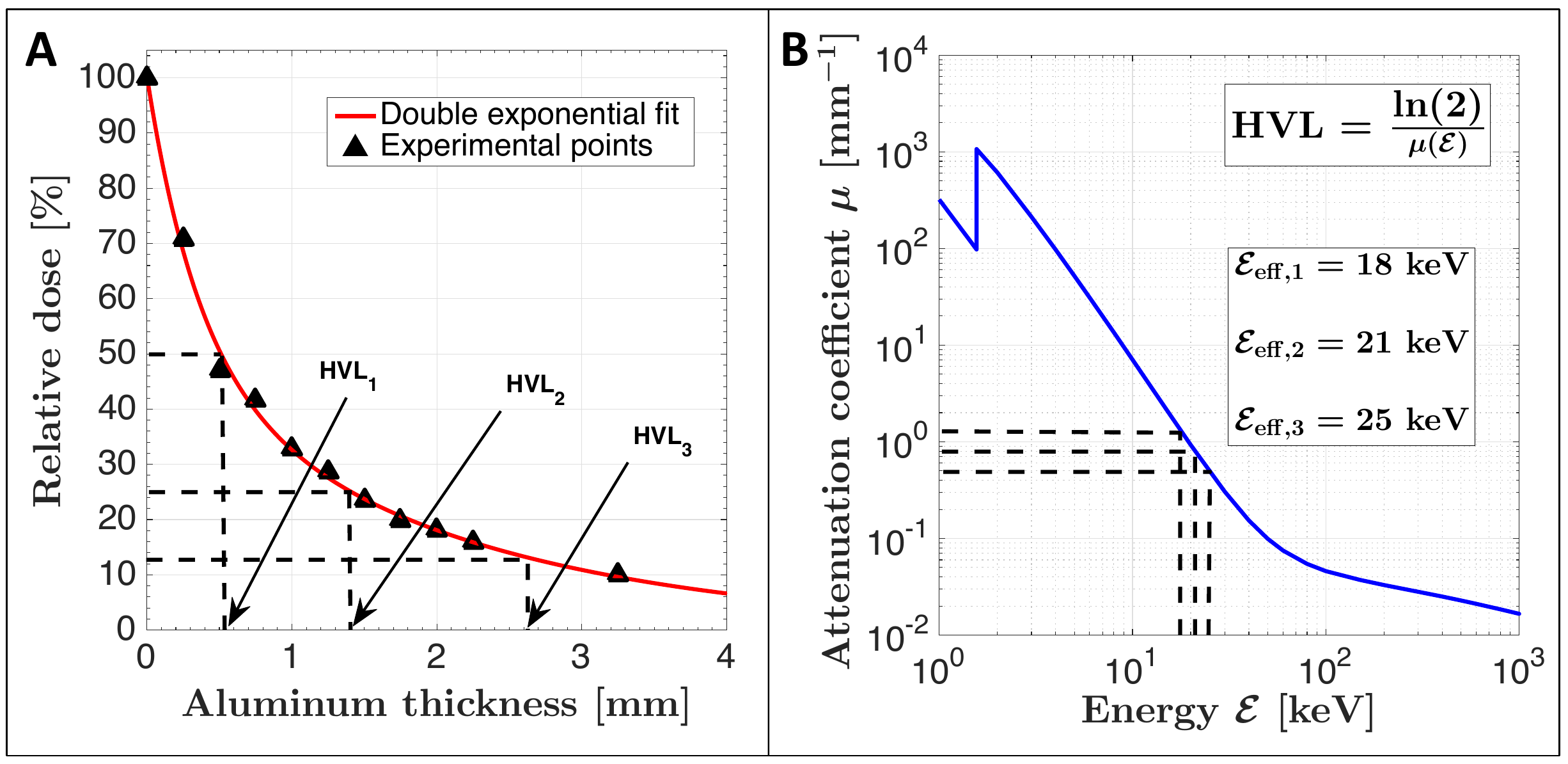}
\caption{Effective X-ray photon energy $\mathcal{E}_\textrm{eff}$ evaluation with HVL measurements using detector D2 at 5 m from the source (no electrons, only X-rays). (A) Aluminum attenuation of photon beam, for which data points (black triangles) were fitted with a double exponential (\textit{i.e.}, two exponential terms, full red line), fitting with $R^2 > 0.99$. Three HVLs were measured: HVL$_1$ = 0.51 mm, HVL$_2$ = 0.90 mm and HVL$_3$ = 1.31 mm. (B) Mean photon energy of the spectrum ($\mathcal{E}_\textrm{eff,1} = 18$ keV, $\mathcal{E}_\textrm{eff,2} = 21$ keV and $\mathcal{E}_\textrm{eff,3} = 25$ keV) for the three retrieved HVLs using the mass attenuation coefficient tables of aluminum from the NIST-XCOM database\cite{nist}.}
\label{fig4}
\end{figure*}

\noindent Given the current laser parameters, the electron acceleration is caused by the relativistic ponderomotive force and not by the process of LWFA\cite{Lu2007,Faure2019,Woodbury2018}, especially because of the tight focusing geometry, gas density and pulse duration of the present work. This is validated by PIC simulations using the 2D3V PICLS code\cite{picls} (see \textit{Supplement 1}). Full 3D PIC simulations, including a tightly-focused EM field model, will further improve the model and be the subject of future studies. In-depth discussion on the mechanism is presented later in the manuscript. At the same laser facility, Veltri et al.\cite{Veltri2017} produced sub-relativistic electrons with energies of a few tens of keV to generate in-air plasma-induced luminescence. The MeV-range electron energies observed in our work are the result of improvements in the laser beamline and the use of a higher numerical aperture focusing optic which increased the intensity at the focus.\\

\noindent Figure \ref{fig4} shows the data regarding the Half-Value Layer (HVL) determination of the photon beam. The dose was measured with detector D2 at 5 m from the source to ensure the measurement of only photons. Aluminum plates of different thicknesses were sequentially placed in front of the detector to filter the photon beam. The data points (black triangles) were then normalized to the maximum dose (\textit{i.e.}, no aluminum case) and are shown in Figure \ref{fig4}.A.  Using a double exponential fit (full red line) to model low and high energy photon populations, it was possible to retrieve three HVLs with values of HVL$_1$ = 0.51 mm, HVL$_2$ = 0.90 mm and HVL$_3$ = 1.31 mm. The excellent fitting ($R^2>0.99$) of the double exponential function is characteristic of a keV polychromatic photon beam. Using the Beer-Lambert law, as shown on Figure \ref{fig4}.B, the mass attenuation coefficient of aluminum was obtained from the NIST-XCOM database\cite{nist} and interpolated to retrieve the effective photon energies of the spectra at different HVL filtration thicknesses, leading to $\mathcal{E}_\textrm{eff,1} = 18$ keV, $\mathcal{E}_\textrm{eff,2} = 21$ keV and $\mathcal{E}_\textrm{eff,3} = 25$ keV for HVL$_1$,  HVL$_2$ and HVL$_3$, respectively. The observed increase in the effective photon energy is characteristic of the so-called beam hardening effect where lower energy photons are preferentially absorbed due to the $1/\mathcal{E}^3$ dependence of the photoelectric effect, which is the dominant interaction type in this energy range. Even though the effective photon energy is measured in the tens of keV, the presence of photons in the hundreds of keV is also expected. Note that the effective energy of an X-ray spectrum is the energy equivalent to a monochromatic photon beam with similar attenuation characteristics (\textit{i.e.}, same HVL, see \textit{Supplement 1} for the definition).

\begin{table}
    \centering
    \caption{Radiation doses reported from the two OSLD area dosimeters from \textit{Health Canada}, accumulated over the course of the experiment. The dosimetric quantity H$_\textrm{p}$(0.07) reports the "surface dose" whereas H$_\textrm{p}$(10) reports the "body dose", equivalent to depths of 0.07 and 10 mm in water, respectively.}
\begin{tabular}{c|ccc}
\hline
\hline
   Dosimeter  & Location & H$_\textrm{p}$(0.07) & H$_\textrm{p}$(10) \\
    & $(r,\theta)$ & [mGy] & [mGy] \\
    \hline
    OSLD1 & (0.7 m, $90^\circ$) & 233.48 & 36.57   \\
   OSLD2 & (2 m, $45^\circ$) & 1238.30 & 9.91  \\
    \hline 
    \hline
\end{tabular}
    \label{tab:OSLD}
\end{table}

\noindent Additionally, the surface (\textit{i.e.}, mostly electrons) and body (\textit{i.e.}, mostly X-rays) doses are reported for the two area dosimeters OSLD1 and OSLD2 from \textit{Health Canada}, as seen in Table \ref{tab:OSLD}. Dose was accumulated over the entire course of the experiment which consisted of nearly 100 acquisitions of 1 minute. OSLD2 reported a surface dose 5.3$\times$ higher than OSLD1, even though it is located farther away from the source (expected to be $(2/0.7)^2 \approx 8\times$ lower from ISL), consistent with the measured forward directionality of the electron beam. Regarding the body dose, OSLD1 reported a dose 3.7$\times$ higher than OSLD2 because of the more isotropic emission of X-ray photons.\\

\section{Discussion}

\subsection{Acceleration mechanism and source optimization}
\noindent

Diffraction-limited focal spots are typically not achieved when focusing a mJ-class, ultrashort laser in ambient air due to both wavefront distortions from the strong Kerr effect and plasma generation that destroy the integrity of the focused laser beam. The $B$-integral quantifying the non-linear phase shift can be expressed as\cite{Ettoumi2010}:
\begin{equation} \label{kerr}
\Delta\phi_\textrm{Kerr} = \sum_{\textrm{N} \in \mathbb{N}^+} B_\textrm{2N} = \frac{2\pi}{\lambda_0} \sum_{\textrm{N} \in \mathbb{N}^+}   \displaystyle\int n_\textrm{2N}(\lambda_0) \, \left[I(z)\right]^\textrm{N} \, \D z
\end{equation}
where N is the N$^\textrm{th}$-term of the Kerr effect, $n_\textrm{2N}$ is the non-linear refractive index and $I(z)$ is the laser intensity along the propagation axis $z$. The $B$-integral needs to be minimized in order to produce a near diffraction-limited focus and the accompanying higher peak intensity. From Equation \ref{kerr}, the wavelength as well as the non-linear refractive index of the medium are critical to determine the amount of phase shift in the laser beam. In gases, the non-linear refractive index $n_\textrm{2N}$ typically decreases with increasing wavelength\cite{Tarazkar2014,Bache2012}, and dramatically decreases for higher ionization states which further limits the $B$-integral during focusing\cite{Tarazkar2016} after the first ionization level. Using the Ammosov-Delone-Krainov (ADK) model for tunnel ionization\cite{Yudin2001,nist3}, the first ionization of air molecules (N$_2$ and O$_2$, \textit{i.e.} the main air constituents) was estimated to occur in the range of $(1-2)\times10^{14}$ W/cm$^2$, for which the cumulative Kerr phase shift was calculated to be $\vert\Delta\phi_\textrm{Kerr}\vert<0.897$ rad ($<\lambda_0/7$) using four terms up to $n_8$\cite{Loriot2009}, hence yielding minor aberrations (see \textit{Supplement 1} for more information). This is due to the use of a relatively long central wavelength that reduces the $B$-integral through the $1/\lambda_0$ dependence, and the tight-focusing geometry that distributes the incident laser energy over a greater focusing solid angle. The ADK model also estimated the maximum ionization state for nitrogen and oxygen to be 5+ and 6+, respectively, at the calculated peak intensity of $1\times10^{19}$ W/cm$^2$. This leads to a calculated electron density of $n_\textrm{e}=2.65\times10^{20}$ cm$^{-3}$ within the plasma which is 23\% below the critical density ($n_\textrm{e} = 0.77n_\textrm{c}$) of $n_\textrm{c}=\varepsilon_0 m_\textrm{e}\omega_0^2/e^2=3.44\times10^{20}$ cm$^{-3}$ at $\lambda_0=1.8$ $\mu$m. The plasma is underdense and therefore transmissive to the laser, allowing a high laser intensity to be achieved without significant plasma defocusing effects.\\

\noindent While most ionization events occur during the leading edge of the laser pulse when the intensity climbs above $10^{14}$ W/cm$^2$, the free electrons produced by ionization of the gas are then driven by the relativistic ponderomotive force when the peak laser intensity reaches $I_0 > 4\times10^{17}$ W/cm$^2$, which is the relativistic intensity threshold ($a_0=1$) calculated for $\lambda_0 = 1.8$ $\mu$m. For intensities above the threshold, electrons oscillate along the $\mathbf{E}$-field polarization at velocities near the speed of light and, combined with the significant $\mathbf{B}$-field that is characteristic of the relativistic regime, feel a strong $\mathbf{v \times B}$ term from the Lorentz force. This accelerates the electrons in the forward direction explaining the high directionality of the electron beam (see Figure \ref{fig3}.A). The calculated electron energy due to the ponderomotive force is also in agreement with the estimated maximum electron energy. This mechanism for the acceleration is further validated by 2D PIC simulations that show the conically-shaped electron beam, the proper range of observed electron energies and the absence of an efficiently accelerating plasma wave for LWFA. The strong laser-matter interaction stems from the near-critical density plasma that enables high laser energy absorption by the electrons (see \textit{Supplement 1} for further details). This efficient coupling explains the high flux of energetic electrons and the resulting large dose rate generated with only mJ of input laser energy.\\

\noindent Further optimization of the source is planned to increase both the electron energy and the dose rate. Higher pulse energies will increase the number of accelerated electrons from a larger ionization volume and increase the ponderomotive energy associated with the higher laser intensities, as observed with the non-linear dose-energy scaling. The use of longer central wavelengths in the mid-IR will reduce the effect of optical beam aberrations as well as further limit the $B$-integral contribution of non-linear effects in air. Increasing the wavelength will decrease the peak intensity as $I_0 \propto 1/\lambda_0^2$ but will keep the ponderomotive energy the same since $\mathcal{E}_{\textrm{e}}^\textrm{pond} \sim I_0\,\lambda_0^2$. The net effect will be an increased ionization volume ($V_\textrm{focal} \propto \lambda_0^3$ in tight focusing) and consequentially a higher measured dose. The influence of a longer wavelength at $\lambda_0=3.9$ $\mu$m on the ponderomotive scaling was also shown in the work by Weisshaupt \textit{et al.}\cite{Weisshaupt2014}, providing a 25$\times$ higher X-ray flux using a $K_\alpha$-based source with solid targets, compared to a central wavelength of $\lambda_0=800$ nm at the same laser intensity $I_0$. An upper limit is foreseen in air around $\lambda_\textrm{c} = (2\pi c/e)\sqrt{\varepsilon_0m_\textrm{e}/n_\textrm{e}} \approx 2.1$ $\mu$m when the critical density decreases to a level similar to the plasma density (\textit{i.e.}, $n_\textrm{c}\approx n_\textrm{e}$) and hinders the propagation of the pulse closer to the focus. This limitation can vary if the gas type and pressure are properly chosen in order to reduce the plasma density. Moreover, the onset of the Relativistic Self-Induced Transparency (RSIT)\cite{Vshivkov1998,Cattani2000,Palaniyappan2012}, into which $n_\textrm{e}\to n_\textrm{e}/\gamma$ at relativistic laser intensities ($a_0>1$), can loosen the constraint on the critical density and further enable the use of longer wavelengths at high intensities for radiation generation. The mechanism is rather complex but a simple estimation with $a_0=5$ (\textit{i.e.} $I_0\approx1.1\times10^{19}$ W/cm$^2$) would bring the 2.1 $\mu$m limit in air to $\lambda^\textrm{RSIT}_\textrm{c}=\lambda_\textrm{c}\times \sqrt{\gamma} =\lambda_\textrm{c}\left(1+a_0^2/2\right)^{1/4}= 4.0$ $\mu$m.\\

\noindent The presented experimental implementation to generate an electron beam is much simpler compared to other laser-based sources such as LWFA since no vacuum setup is required nor an increased gas density through the use of a gas jet. LWFA acceleration also requires finer tuning for a particular set of laser parameters to reach a well-defined acceleration regime, along with a synchronization of laser beam with the gas jet. Our technique can scale to much higher repetition rates and increased pulse energies to enhance both the rate of electron production (\textit{i.e.}, dose) and the electron energy. Through further optimization, it is foreseen that in-air, laser-based high dose-rate sources of this type will provide a critical platform for ionizing radiation applications. Furthermore, we emphasize the great potential of this technique for studying the FLASH effect in radiobiology (see section \ref{flash_sec}) due to its ease of implementation and ability to not only provide a high instantaneous dose rate but also a very high average dose rate.  Future work will also investigate the measurement of the radiation pulse duration in order to characterize the instantaneous dose rate.\\

\subsection{Radiation safety in laser facilities}
\noindent

In light of the high doses reported in this work, along with the simple design of the setup, it is important to delineate the possible consequences. In Canada, the dose limits are determined by the Canadian Nuclear Safety Commission (CNSC), which sets a limit of 1 mGy (for electrons and photons) per year for the general public\cite{cnsc}. However, in the radiation protection community, there is a consensus to use a more conservative As Low As Reasonably Achievable (ALARA) limit of 0.1 mGy per year. This ensures a long-term radiation safety environment for personnel who frequently work close to ionizing radiation. It is possible to note from Figure \ref{fig2}.A that at the lowest laser pulse energies and several meters away from the source, the CNSC limit of 1 mGy is reached after a few hours of exposure. A typical researcher in the laboratory can easily spend several hours near the focused beam, potentially suffering serious exposure to ionizing radiation in a surprisingly short time. At a typical standing distance of one meter from the focusing element, this dose limit is reached within one minute at 1.2 mJ and under one second at 2.8 mJ. This leads to serious concerns with the physical proximity of a person to this type of setup; that is when tightly focusing in air with a mJ-class femtosecond laser. Continued development and commercialization of the next generation of multi-mJ lasers operating at kHz repetition rate and above will rapidly escalate the potential for laboratory personnel to suffer even higher dose exposure. In particular, the repetition rate can be increased by up to three orders of magnitude using either Ytterbium-pumped Optical Parametric CPA (OPCPA)\cite{Zou2021} or Thulium-doped fiber lasers\cite{Gaida2021,Gaida2021_2}. We strongly advise that researchers, working with lasers under tight focusing conditions, exercise extreme caution and maintain proper radiation safety measures and training. Regarding the area dosimeters, OSLD1 reported a dose 2330$\times$ higher than the ALARA dose limit, whereas OSLD2 was 12380$\times$ higher, for the duration of just one single experimental campaign. This demonstrates the high degree of danger that such an experimental configuration can present, and the required attention to the necessary precautions. Finally, we recommend to the community who use ultrashort mJ-class lasers to review their laboratory guidelines and procedures for radiation protection.\\

\subsection{Application to FLASH-RT}\label{flash_sec}
\noindent

The FLASH effect\cite{Vozenin2019,Wilson2020,Lin2021} is a radiobiological outcome present at very high dose rates in ultrafast time frames that, when compared to conventional radiation therapy (RT) treatments, promises to deliver fewer normal tissue complications for the same biological damage to malignant cells. This topic generated excitement in the RT community in 2014 after successful demonstrations in mice\cite{Favaudon2014, MontayGruel2017}, cats and mini-pigs\cite{Vozenin2018}. These noticeable FLASH-RT effects were observed for average dose rates of $\dot{\overline{D}} = D_\textrm{pulse} f_\textrm{rep} \gtrsim 40$ Gy/s and maximized above $\sim$$100$ Gy/s, where $D_\textrm{pulse}$ is the dose per pulse and $f_\textrm{rep}$ is the repetition rate. Table \ref{tab:flash} shows a comparison of the dose rate characteristics of laser-based electron beams relative to current electron FLASH-RT sources and the continuous beam radiation of conventional RT (0.1 Gy/s). Even though it is not practical to precisely compare these numbers as beam parameters and experimental conditions are different, it is useful to note the advantages of the different techniques. In this work, we estimate that the on-axis dose per pulse at the same endpoint of 3 mJ and 0.4 m was 0.36 mGy which is $3\times$ lower than the work of Cavallone \textit{et al.}\cite{Cavallone2021}. This difference is explained by the high density supersonic gas jets used\cite{Gustas2018}, the longer focal length of their focusing optic (\textit{i.e.}, larger focal volume), and the shorter pulse duration used in the LWFA process. It is interesting to note, though, that the simplicity of our setup permits access to the radiation at much shorter distances from the source. For example, at 0.1 m we estimate that the on-axis dose per pulse is 3.5$\times$ greater at 3.8 mGy. It is important to point out, however, that both experiments can be repeated with higher repetition rate lasers which would increase the average dose rates to values relevant for FLASH-RT, thereby confirming their potential.\\

\noindent Due to the nature of our electron acceleration mechanism, the instantaneous dose rate $\dot{D} = D_\textrm{pulse}/\tau$ obtained in this work is estimated to be as high as $10^9$ Gy/s for picosecond pulse durations at few millimeters away from the interaction volume. This is orders of magnitude above most electron FLASH-RT sources except for large-scale facilities, which typically have more electrons per pulse spread over much longer time scales\cite{Sampayan2021}. In the recent work of Vozenin \textit{et al.}\cite{Vozenin2020}, the 40 Gy/s average dose rate definition of FLASH-RT has proven to be over simplistic. Accumulated evidence rather points towards the instantaneous dose rate and overall irradiation time as the critical parameters. This highlights ultrafast laser-based radiation as prominent sources to investigate the FLASH effect at ultra-high instantaneous dose rates. The average dose rate requirements of FLASH-RT may be lowered if sufficiently high instantaneous dose rate is reached. In Table \ref{tab:flash}, the large spread of dose rates reported with current $e^-$ FLASH-RT sources range from table-top systems to large-scale particle accelerator facilities. As noted in the work of Bourhis \textit{et al.}\cite{Bourhis2019}, even if room-sized systems such as modified LINACs are more affordable, there are very few available worldwide. Their work also noted that a logical first step towards the clinical translation of FLASH-RT is to use low-energy electrons of a few MeV in pre-clinical conditions as a proof of concept of the FLASH effect in human patients. Hence, a laser-based electron beam, as presented in our experimental configuration, is an ideal candidate in this matter.\\
 
\noindent Compared to our work, LWFA sources from large-scale J-class laser facilities\cite{Lundh2012,Svendsen2021} can exhibit similar dose per pulse, higher instantaneous dose rates up to $10^{11}$ Gy/s due to the ability to produce femtosecond electron bunch durations, but a much lower average dose rates as a result of the limited repetition rate. Hence, J-class LWFA sources are still orders of magnitude away from reaching the average dose rate requirements of FLASH-RT and are even below conventional RT. However, they are capable of generating hundreds of MeV electron beams for Very High Energy Electron (VHEE) treatments which the few MeV-electrons of mJ-class lasers cannot do.

\begin{table}
    \centering
    \caption{Summary of dose rate characteristics for various electron beam sources. The value $\dot{\overline{D}}$ is the on-axis ($\theta=0^\circ$) average dose rate, $f_\textrm{rep}$ is the repetition rate, $D_\textrm{pulse}$ is the dose per pulse and $\dot{D}$ is the estimated instantaneous dose rate. *The number in parentheses is at 0.4 m from the source and 3 mJ per laser pulse.}
\begin{tabular}{c|cccc}
\hline
\hline
     Source & $\dot{\overline{D}}$ & $f_\textrm{rep}$ & $D_\textrm{pulse}$ & $\dot{D}$ \\
    & [Gy/s] & [Hz] & [mGy] & [Gy/s] \\
    \hline
   This work & 0.38 & $10^2$ & 3.8 (0.36)* & $10^7-10^9$  \\
   mJ-class LWFA\cite{Cavallone2021} & 1.1 & $10^3$ & 1.1 & $10^7$  \\
   J-class LWFA\cite{Lundh2012,Svendsen2021} & 0.03 & $10^1$ & 3 & up to $10^{11}$  \\
   $e^-$ FLASH-RT\cite{Wilson2020,Sampayan2021} & $10^1-10^6$ & $10^1 - 10^6$ &  $10^2 - 10^{4}$ & $10^3 - 10^{10}$  \\
   Conventional RT & 0.1 & --- &  --- & 0.1  \\
    \hline 
    \hline
\end{tabular}
    \label{tab:flash}
\end{table}

\noindent Note that the measured dose in our work was likely under-estimated since the detection range of both detectors is not sensitive to electrons with energies below 100 keV due to their absorption in the detector wall. Ponderomotive electrons will have an energy distribution with some proportion of electrons below this detection threshold. The use of radiochromic films (RCF) for dose measurements is to be considered in subsequent experiments to overcome this problem since they are dose-rate independent for a wide range of dose rates up to 10$^{12}$ Gy/s\cite{Karsch2012,Jaccard2017,Bazalova2021}. Nevertheless, both detectors D1 and D2 were well in their range of dose rate usability for the present experiment\cite{Permattei2000,Kranzer2021}. Due to the ultrafast nature of the laser-driven electron acceleration mechanism, this source is a very promising candidate for characterizing the potential of the FLASH effect.\\

\section{Conclusion}
\noindent

This work reports on the generation of a high dose-rate MeV electron beam produced by direct laser acceleration in ambient air using a mJ-class femtosecond IR laser operated at 100 Hz repetition rate. The electron beam reaches the MeV-level of kinetic energy through the relativistic ponderomotive force from the laser, with the measured beam characteristics further supported by 2D PIC simulations. This is enabled by the low $B$-integral from the use of a 1.8 $\mu$m central wavelength and a tight focusing geometry. The highest dose rate reported in this work of 0.15 Gy/s (0.38 Gy/s estimated on-axis) is several times higher than the conventional dose rates used in clinical radiotherapy for cancer treatments. This dose rate can deliver the yearly personal radiation exposure limit in seconds for someone standing a few meters away from the source. This raises a major safety issue for laboratories using mJ-class lasers under tight focusing geometry. We strongly advise the concerned laser users to take the necessary radiation protection precautions. If these aspects are properly handled, this laser-based ionizing radiation source is promising for studying the FLASH effect in radiobiology with an ultrashort electron beam yielding instantaneous dose rates up to $10^9$ Gy/s. Further work will investigate the scaling of the dose, both numerically and experimentally, by varying a series of parameters including laser wavelength, gas, and pressure. Finally, we are planning cellular and small animal irradiation studies in radio-oncology.\\

\begin{backmatter}
\bmsection{Funding}
The authors acknowledge funding from the NSERC-CRD program, S.V. acknowledges FRQNT for the Postdoctoral Research Scholarship. This research was enabled in part by support provided by Compute Canada.

\bmsection{Acknowledgments}
The authors would like to thank Antoine Laram\'ee for the technical support and Fran\c cois Vidal for the fruitful discussions.

\bmsection{Disclosures}
The authors declare no conflicts of interest.

\bmsection{Data availability} Data underlying the results presented in this paper are not publicly available at this time but may be obtained from the authors upon reasonable request.

\bmsection{Supplemental document} See \textit{Supplement 1} for supporting content.

\end{backmatter}




\end{document}


\title{High Dose-Rate Ionizing Radiation Source from Tight Focusing in Air of a mJ-class Femtosecond Laser: Supplemental Document}

\author{Simon Valli\`eres\authormark{1,2,$\dagger$,*}, Jeffrey Powell\authormark{1,$\dagger$}, Tanner Connell\authormark{3}, Michael Evans\authormark{3}, Sylvain Fourmaux\authormark{1}, St\'ephane Payeur\authormark{1}, Philippe Lassonde\authormark{1}, Fran\c cois Fillion-Gourdeau\authormark{4}, Steve MacLean\authormark{1,2,4} and Fran\c cois L\'egar\'e\authormark{1}}

\address{\authormark{1} INRS-EMT, 1650 blvd. Lionel-Boulet, Varennes, QC, J3X 1P7, Canada\\
\authormark{2} Institute for Quantum Computing, University of Waterloo, Waterloo, ON, N2L 3G1, Canada\\
\authormark{3} Medical Physics Unit, McGill University Health Center, 1001 blvd. D\'ecarie, Montr\'eal, QC, H4A 3J1, Canada\\
\authormark{4} Infinite Potential Laboratories, Waterloo, ON, N2L 0A9, Canada}
\hspace{-3.5mm}\authormark{$\dagger$}\textit{ \small Equivalent first-author contribution}\\
\email{\authormark{*} Corresponding author: simon.vallieres@uwaterloo.ca} 




\section{Tightly-focused electromagnetic fields calculation}
\noindent

We simulated the tightly-focused electromagnetic (EM) fields of the laser with an in-house code, called the \textit{Strattocalculator}\cite{Dumont2017}, that uses the Stratton-Chu integral formulation of EM fields\cite{Stratton1939}. The simulation considers an ideal incident linearly-polarized beam (\textit{i.e.} no aberrations) and the propagation of the fields in vacuum. The input parameters of the beam model distributed 2.8 mJ in 128 spectral components across 400 nm of a super-Gaussian spectrum of order 7 centered at $\lambda_0=1.8$ $\mu$m. This yields a Fourier-limited pulse duration of $\tau_\textrm{FWHM}$ = 12 fs. We used a TEM$_{00}$ transverse profile with a beamsize of $w_\textrm{FWHM}$ = 10.5 mm, incident on a on-axis parabolic reflector with outer diameter of $D=25.4$ mm and a focal length of $f=6.35$ mm (\textit{i.e.} parabola numerical aperture of NA = 1,  effective beam numerical aperture of NA = 0.94 at the beamwaist of $w_0$ defined at 1/$e^2$ of the maximum intensity). The calculation time is of 10 hours on 800 cores.

\begin{figure*}[h!]
\centering\includegraphics[width=320pt]{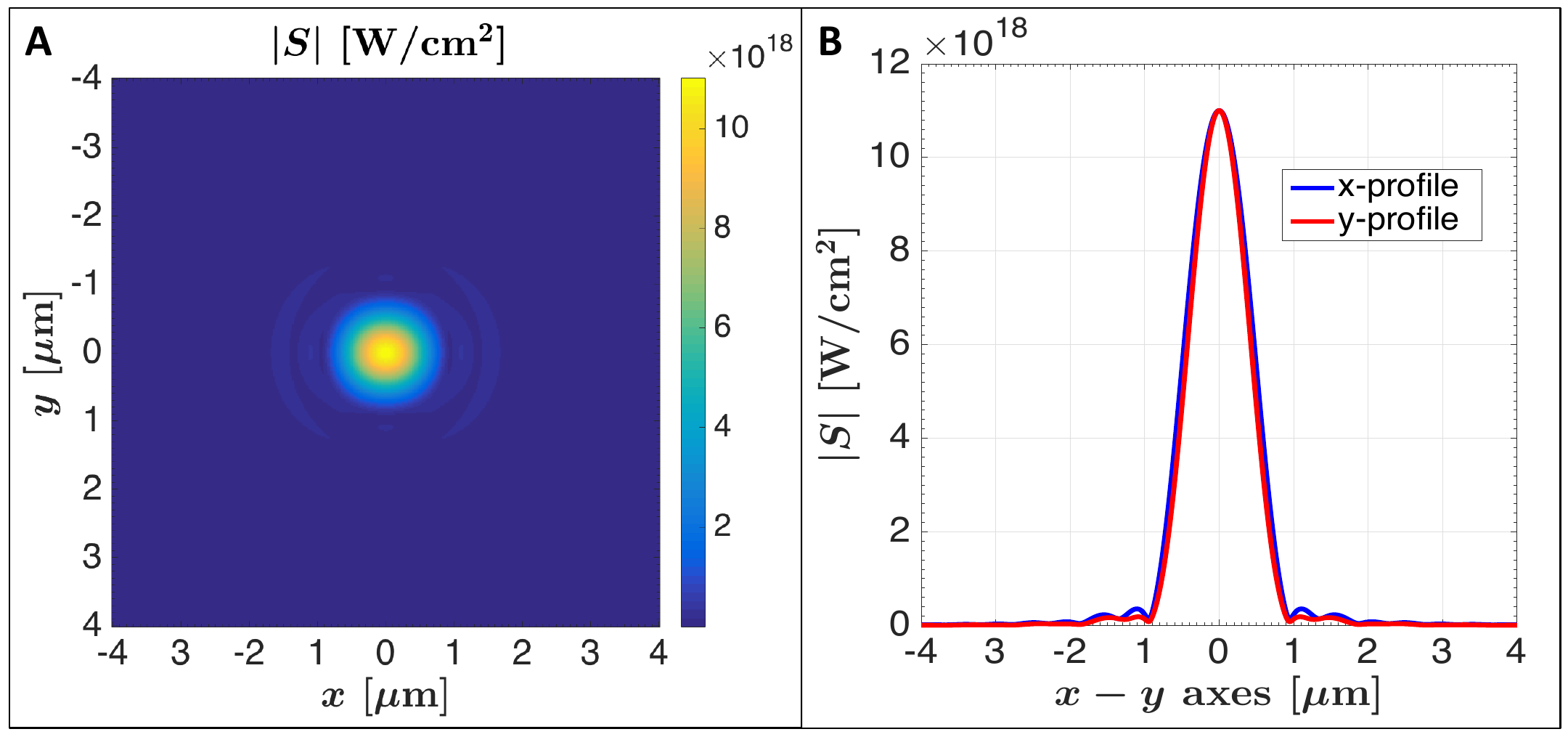}
\caption{Tightly-focused EM fields calculation using the \textit{Strattocalculator}. (A) Norm of the Poynting vector $\vert\boldsymbol{S}\vert$ in the transverse plane. The beam is linearly polarized along the $x$-axis and the peak intensity reaches around $1\times10^{19}$ W/cm$^2$. (B) Intensity profiles of $\vert\boldsymbol{S}\vert$ along both $x$ and $y$ axes, yielding a focal spot size at FWHM of $w_\textrm{FWHM}$ = 1.0 $\mu$m.}
\label{strattofig}
\end{figure*}

\noindent Figure \ref{strattofig} shows the norm of the Poynting vector $\vert\boldsymbol{S}\vert = \vert\boldsymbol{E}\times\boldsymbol{B}\vert/\mu_0$ in the transverse $xy$-plane at the focus position $z=0$, where $\boldsymbol{E}$ is the electric field, $\boldsymbol{B}$ is the magnetic field and $\mu_0$ is the vacuum permeability. Note in Figure \ref{strattofig}.A that the peak intensity reaches $I_0 = \vert\boldsymbol{S}\vert_0 = 1\times10^{19}$ W/cm$^2$ and exhibits a diffraction-limited focal spot size of $w_\textrm{FWHM}$ = 1.0 $\mu$m, as observed in Figure \ref{strattofig}.B. This calculation sets an upper bound to the peak intensity reached in the experiment, as a larger spot size is expected due to non-linear effects and plasma generation due to focusing in air which can distort the wavefront, and reduce the peak intensity.\\

\section{ADK model for tunnel ionization}
\noindent

We used the Ammosov-Delone-Krainov (ADK) model to evaluate the ionization probability for strong-field tunnel ionization, following the methodology described in the work of Yudin \textit{et al.}\cite{Yudin2001}. Assuming an ionic species is ionized from a state with ion charge of $Z$ to another species with a charge of $Z+1$, the ionization rate is given, in atomic units, by the following expression:

\begin{equation}
\Gamma(E,Z) = \frac{2^{2n^{*} + 1}}{\Gamma(2n^{*} +1)} (2\ell + 1) \exp\left[ - \frac{\Delta}{3} + (2n^{*} - 1) \ln (\Delta) \right]
\end{equation}

\noindent where $n^{*} = (Z+1)/\sqrt{2I_{p}(Z)}$, $I_{p}(Z)$ is the ionization potential of ionic species $Z$, $\ell$ is the angular quantum number and $\Delta = 2[2I_{p}(Z)]^{\frac{3}{2}}/E(t)$, with $E(t)$ being the electric field strength. The unit quantities to convert to SI units are defined in Table \ref{tab:unit}, and the values of the ionization potentials are taken from the NIST-ASD database\cite{nist3}.

\begin{table}[h!]
	\centering
	\caption{Unit quantities to transform from/to atomic units.}
	\begin{tabular}{l c} 
		\hline
		Physical quantities & Unit quantities \\
		\hline
		Rate & $4.134 \times 10^{16}$ Hz \\
		Field strength & $0.514 224 \times 10^{12} $ V/m \\
		Ionization potential & 27.2116 eV \\
		\hline
	\end{tabular}
\label{tab:unit}
\end{table}

\noindent Once the rate is evaluated, the ionization probability is given by:

\begin{equation}
P(t) = 1 - e^{-\int_{-\infty}^{t}\Gamma[E(t'),Z] \, \D t'}
\end{equation}

\noindent To estimate the ionization probability in the experiment, we assume a simple Gaussian temporal envelope of the form:

\begin{equation}
E(t) = E_{0} e^{-2 \ln(2) \frac{t^{2}}{\tau^2_\textrm{FWHM}}} \cos(\omega_0 t + \phi)
\end{equation}

\begin{figure*}[h!]
\centering\includegraphics[width=320pt]{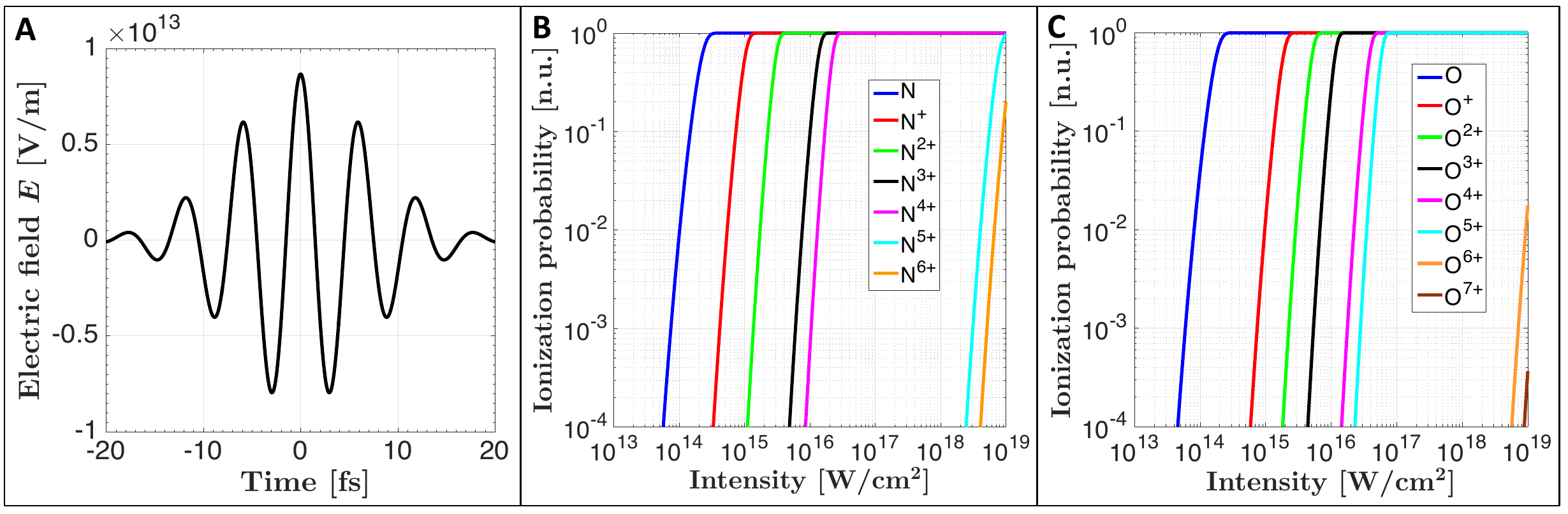}
\caption{ADK model for tunnel ionization. (A) Electric field strength form used in the calculation. The ionization probability resulting from different peak intensity is obtained for different ionic species of (B) nitrogen and (C) oxygen.}
\label{adkfig}
\end{figure*}

\noindent where $E_{0}$ is the maximum field strength, $\tau_\textrm{FWHM}$ is the Full-Width-Half-Maximum (FWHM) pulse duration (in intensity), $\omega_0$ is the central angular frequency and $\phi$ is the Carrier-Envelope Phase (CEP). The field strength is fixed from the peak intensity $I_0$ as $E_{0} = \sqrt{2I_0/c\varepsilon_{0}}$. The data for nitrogen and oxygen are displayed in Figure \ref{adkfig} for $I_0 = 1 \times 10^{19}$ W/cm$^{2}$, $\tau_\textrm{FWHM} = 12$ fs, $\phi = 0$ and $\lambda_0 = 1.8$ $ \mu$m. We can note from Figure \ref{adkfig}.B-C that the first ionization of both N and O occurs in the range of $I_0 = [1-2] \times 10^{14}$ W/cm$^{2}$, where the probability reaches $P(t)>0.5$. For the highest peak intensity of $I_0 = 1 \times 10^{19}$ W/cm$^{2}$, the highest ionization state reaches 5+ and 6+ for nitrogen and oxygen, respectively, which yields a maximum electron density in air of $n_\textrm{e}=2.65\times10^{20}$ cm$^{-3}$, corresponding to $n_\textrm{e} = 0.77n_\textrm{c}$ for $\lambda_0$ = 1.8 $\mu$m.\\

\section{Higher-order Kerr effect $B$-integral evaluation}
\noindent

\begin{figure*}[h!]
\centering\includegraphics[width=280pt]{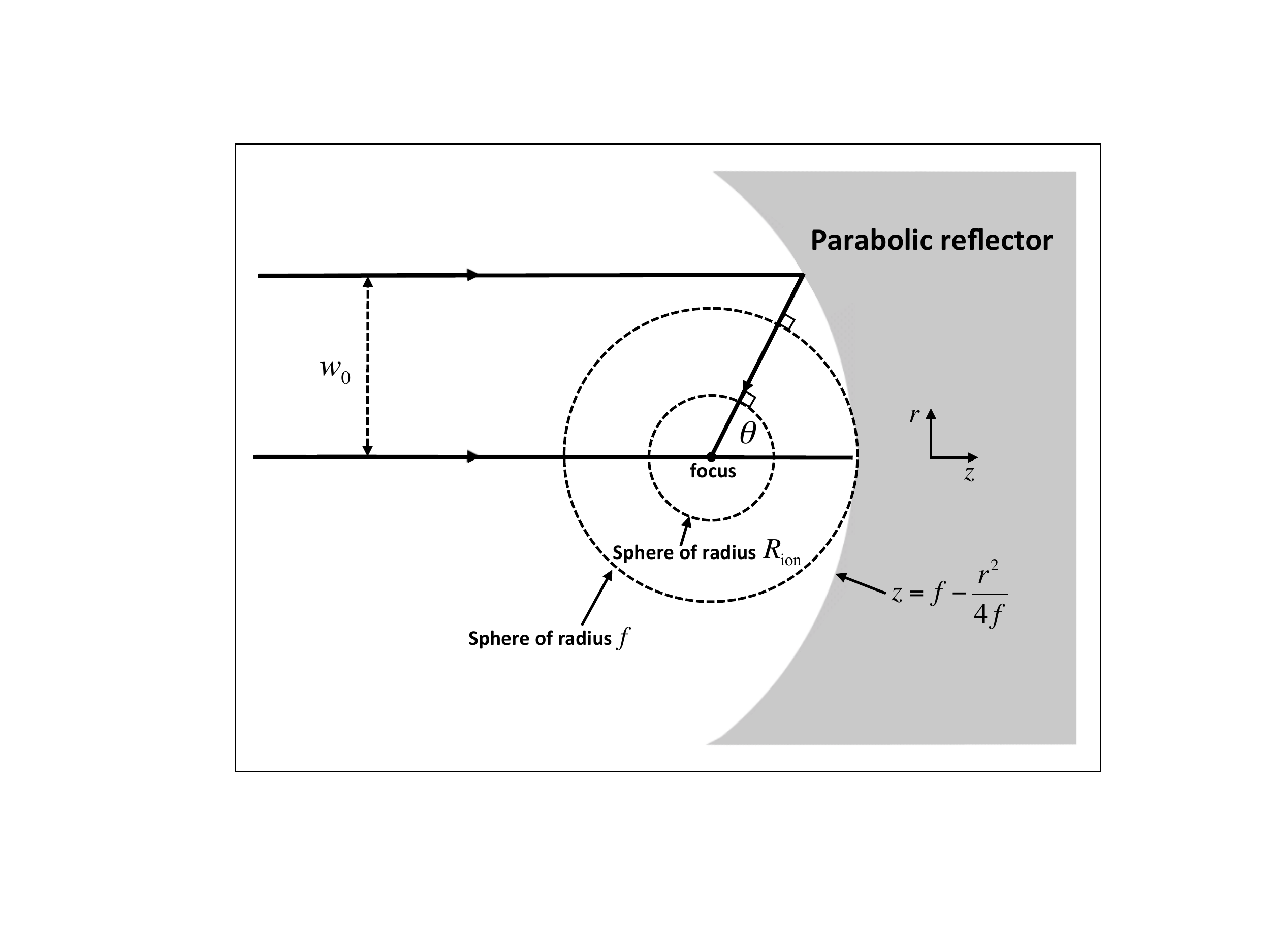}
\caption{Schematic diagram of the focusing spheres after the reflection on the parabolic reflector.}
\label{Bfig}
\end{figure*}

The $B$-integral quantifying the non-linear Kerr effect phase shift $\Delta\phi_\textrm{Kerr}$ can be expressed as\cite{Ettoumi2010}:

\begin{equation} \label{kerr}
\Delta\phi_\textrm{Kerr} = \sum_{\textrm{N} \in \mathbb{N}^+} B_\textrm{2N} = \frac{2\pi}{\lambda_0} \sum_{\textrm{N} \in \mathbb{N}^+} \displaystyle\int n_\textrm{2N}(\lambda_0) \, \left[I(z)\right]^\textrm{N} \, \D z
\end{equation}

\noindent where N is the N$^\textrm{th}$-term of the Kerr effect, $n_\textrm{2N}$ is the non-linear refractive index and $I(z)$ is the laser intensity along the propagation axis $z$. In order to evaluate equation \eqref{kerr}, we first find the focusing angle $\theta$ at the beamwaist $w_0 = w_\textrm{FWHM}/\sqrt{2\ln(2)} = 8.91$ mm using the parabolic equation of the reflector as:

\begin{equation}
z(w_0)=f - w_0^2/4f \hspace{1cm} \implies \hspace{1cm}  \theta = \arctan\left[\frac{w_0}{f - w_0^2/4f} \right] \approx 70^\circ
\end{equation}

\noindent where $f = 6.35$ mm. The angle $\theta$ is delimiting the focusing cone of light with its apex located at the geometrical focus at $z=0$, as shown in Figure \ref{Bfig}. During the beam focusing after the reflection on the parabola, the laser intensity will rise as the beam area reduces while the beam is traveling closer to the focus along the longitudinal axis $z$. We can then find the solid angle $\Omega$ of the light cone using:

\begin{equation}
\Omega = 2\pi(1-\cos\theta) = \frac{A}{R^2}
\end{equation}

\noindent where $A$ is the segment area of the sphere of radius $R$ centered at the focus. Assuming a uniform beam profile (top-hat), the intensity follows $I = \frac{dP}{dA} = P_0/A$, where $P_0$ is the peak laser power. To find $P_0$ from the laser pulse energy of $\mathcal{E}_\textrm{L}$=2.8 mJ, we assume a Gaussian temporal shape of the form:

\begin{equation}
\mathcal{E}_\textrm{L} = \displaystyle\int\limits_{-\infty}^{\infty} \, P_0 e^{-4 \ln(2) \frac{t^{2}}{\tau^2_\textrm{FWHM}}}  \, \D t \hspace{1cm} \implies \hspace{1cm} P_0 = \sqrt{\frac{4\ln(2)}{\pi}} \frac{\mathcal{E}_\textrm{L}}{\tau_\textrm{FWHM}} = 0.22 \textrm{ TW}
\end{equation}

\noindent We can then find the following expression for the peak intensity on-axis ($\theta=0^\circ$) along $z=R\cos\theta$:

\begin{equation}
I(z) = \frac{P_0}{A} =  \frac{P_0}{\Omega z^2}
\end{equation}

\noindent Assuming negligible $B$-integral accumulation before the reflection of the beam on the focusing optic, we therefore perform the integration from the sphere of radius $f$ down to the radius where the first ionization occurs at $R_\textrm{ion}$. Considering the first ionization of nitrogen and oxygen in air takes place in the range of $I_\textrm{ion} = (1-2)\times 10^{14}$ W/cm$^{2}$, we can find:

\begin{equation}
R_\textrm{ion} = \sqrt{\frac{P_0}{\Omega I_\textrm{ion}}} \approx 230 \textrm{ }\mu\textrm{m}
\end{equation}

\noindent Replacing the variables in equation \eqref{kerr}, we obtain:

\begin{equation} \label{kerr2}
\Delta\phi_\textrm{Kerr} = \frac{2\pi}{\lambda_0} \sum_{\textrm{N} \in \mathbb{N}^+} \displaystyle\int\limits^{f}_{R_\textrm{ion}} n_\textrm{2N} \, \left( \frac{P_0}{\Omega z^2}\right)^\textrm{N} \, \D z = \frac{2\pi}{\lambda_0} \sum_{\textrm{N} \in \mathbb{N}^+} n_\textrm{2N} \left(\frac{P_0}{\Omega}\right)^\textrm{N} \displaystyle\int\limits^{f}_{R_\textrm{ion}} \, z^\textrm{-2N} \, \D z
\end{equation}

\noindent which has the following solution:

\begin{equation} \label{kerr3}
\Delta\phi_\textrm{Kerr} = \frac{2\pi}{\lambda_0} \sum_{\textrm{N} \in \mathbb{N}^+} n_\textrm{2N} \left(\frac{P_0}{\Omega}\right)^\textrm{N} \left(\frac{f^{1-2\textrm{N}}-R_\textrm{ion}^{1-2\textrm{N}}}{1-2\textrm{N}} \right)
\end{equation}

\begin{table}
    \centering
    \caption{High-order $B$-integral evaluation for the first four terms up to $n_8$ and integrated to a maximum intensities of $(1-2)\times10^{14}$ W/cm$^2$.}
\begin{tabular}{c|cccc|c}
\hline
\hline
  $I_\textrm{ion}$ & $B_2$ & $B_4$ & $B_6$ & $B_8$ & $\Delta\phi_\textrm{Kerr}$ \\
  \hspace{0.1mm} [W/cm$^2$] & [rad] & [rad] & [rad] & [rad] & [rad] \\
    \hline
   $1\times10^{14}$ & 0.0093 & -0.004 & 0.0338 & -0.0919 & -0.0528 ($\lambda_0/120$) \\
    $2\times10^{14}$ & 0.0133 & -0.0114 & 0.1910 & -0.0919 & -0.8466 ($\lambda_0/7.4$) \\
    \hline 
    \hline
\end{tabular}
    \label{tab:kerr}
\end{table}

\noindent In Table \ref{tab:kerr} we list the calculated $B$-integral terms using the non-linear refractive indices for air, which are available in the work of Loriot \textit{et al.}\cite{Loriot2010} up to $n_8$. It is possible to observe that $\vert\Delta\phi_\textrm{Kerr}\vert<0.897$ rad for both intensities, which corresponds to beam aberrations of $\lambda_0/7$ in amplitude. We can note from equation \eqref{kerr3} that the low $B$-integral accumulation stems from the combination of both the long central wavelength of $\lambda_0 = 1.8$ $\mu$m and the large focusing solid angle $\Omega$ from the very high numerical aperture of the focusing optic (NA$_\textrm{parabola}$ $\approx 1$, effective NA = 0.94 at the beamwaist $w_0$). Hence, we expect that accumulated aberrations from non-linear effects are low before the first ionization, and that their accumulation is further limited for higher ionization states, as shown in the work of Tarazkar \textit{et al.}\cite{Tarazkar2016}.\\

\section{2D Particle-In-Cell simulations}
\noindent

\begin{figure*}[h!]
\centering\includegraphics[width=320pt]{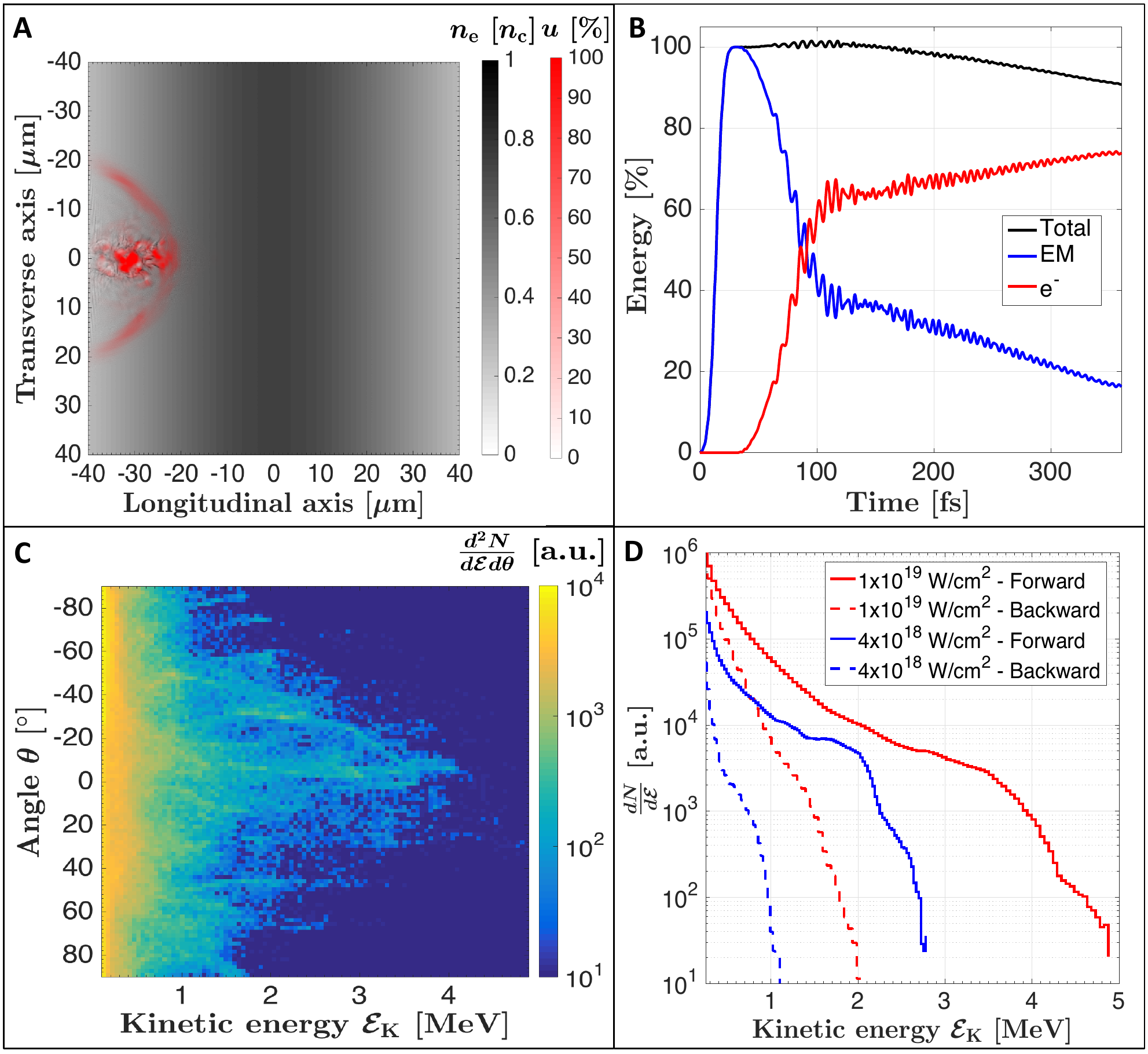}
\caption{Two-dimensional PIC simulation using the 2D3V PICLS code\cite{picls} to model the high-intensity interaction with air. (A) Snapshot of the electron density $n_\textrm{e}$ (grayscale) and the electromagnetic energy density $u$ (red), at $t=120$ fs in the early part of the interaction. A large conical EM shockwave trailing the laser pulse is observed, due to a strong electron current induced from the laser's ponderomotive kick. (B) Total energy (black), EM energy (blue) and electron energy (red) contained in the simulation box as a function of time. All the curves are normalized to the maximum of the EM energy (blue line). Electrons are absorbing as high as 63\% of the laser energy during the initial high-intensity interaction. The total energy goes only up to 101\% of the total EM energy, confirming the good numerical stability of the simulation. (C) Angular energy spectrum $\frac{d^2N}{d\mathcal{E}d\theta}$ of the forward-traveling electrons at $t=180$ fs (after the main interaction). Electrons are forming a cone, centered at 0$^\circ$, exhibiting decreasing divergence with increasing electron energy. (D) Electron spectra at $t=180$ fs for both the high (red) and low (blue) intensities, also for forward (full line) and backward (dashed line) traveling electrons. Sub-figures (A)-(B)-C) are done at $I_0 = 1\times10^{19}$ W/cm$^2$. }
\label{picfig}
\end{figure*}

In order to shed light on the physical origin of the acceleration mechanism, 2D Particle-In-Cell (PIC) simulations were performed using the 2D3V PICLS code\cite{picls}. The beam model used a linearly-polarized Gaussian laser pulse at a central wavelength of $\lambda_0 = 1.8$ $\mu$m, a pulse duration of $\tau_\textrm{FWHM} = 12$ fs and focused down to a Gaussian focal spot size of $w_\textrm{FWHM} = 1$ $\mu$m. We simulated two laser intensities of $1\times10^{19}$ W/cm$^2$ ($a_0 = 4.86$) and $4\times10^{18}$ W/cm$^2$ ($a_0 = 3.08$), where $a_0$ is the normalized amplitude of the vector potential. The simulation grid uses $\Delta x = \Delta y = 120$ nm, $\Delta t = \Delta x/c = 0.4$ fs and runs over 400 fs. The box size is of 1600 transverse cells by 820 longitudinal cells, corresponding to box dimensions of $192\times98$ $\mu$m$^2$ ($95\lambda_0 \times 54 \lambda_0$). Nitrogen ion species were initiated at the $2+$ ionization level with field and collisional ionizations enabled. A Gaussian plasma density profile was used with a standard deviation of 40 $\mu$m ($22\lambda_0$) and maximum density of $n_\textrm{e}=0.75 n_\textrm{c}$, where $ n_\textrm{c}$ is the critical density at $\lambda_0 = 1.8$ $\mu$m. The increasing plasma density serves to simulate the electron injection as the laser intensity rises in the leading edge of the pulse. We used 100 macroparticles per cell for a total of over $10^8$ macroparticles. Each simulation runs for 6 hours on 800 cores. The results of the simulations are presented in Figure \ref{picfig}.\\

\noindent Figure \ref{picfig}.A shows a composite snapshot of the electron density $n_\textrm{e}$ (grayscale) and the electromagnetic energy density $u=\mathlarger{\frac{\varepsilon_0}{2}}\left(\vert\boldsymbol{E}\vert^2 + c^2\vert\boldsymbol{B}\vert^2\right)$ (red), at $t=120$ fs in the early part of the interaction. Note the large EM shockwave, larger than the laser itself, which stems from a strong current of ponderomotive electrons that were driven by the intense laser pulse as it penetrates in the plasma. Figure \ref{picfig}.B shows the total (black), EM (blue) and electron (red) energies contained within the simulation box. It is interesting to note the laser energy coupling to the electrons is efficient, absorbing as high as 63\% of the incident EM energy due to their near-critical density ($n_\textrm{e} \approx n_\textrm{c}$). This strong absorption efficiency led to the observed high electron energies. This is enabled by the long wavelength, which permits a relatively low critical plasma density, and the high intensity reached at the focus, which produces a plasma density near the critical density as deep states of air molecules are ionized. Moreover, note that the total energy (black line in Figure \ref{picfig}.B) goes up to 101\% of the total EM energy (\textit{i.e.} only 1\% of numerical heating), confirming the good numerical stability of the simulation. In Figure \ref{picfig}.C, the angular spectral distribution $\frac{d^2N}{d\mathcal{E}d\theta}$ of the forward-traveling electrons at $t=180$ fs is shown, after the main interaction. It is possible to observe the conical distribution of electrons, centered around the longitudinal axis at 0$^\circ$, that exhibit decreasing conical angles with increasing kinetic energy. The observed noise in the data is due to an increased statistical noise from the 2D histogram, and requires a high number of macroparticles per cell to generate a smooth distribution. Integrating the angular spectral distributions over all angles yields the spectra presented in Figure \ref{picfig}.D. We show the spectra for two intensities ($1\times10^{19}$ W/cm$^2$ in red and $4\times10^{18}$ W/cm$^2$ in blue), for both forward (full lines) and backward (dashed lines) traveling electrons. All the spectra follow a typical Maxwell-Boltzmann distribution where $\mathlarger{\frac{dN}{d\mathcal{E}}} \sim e^{\mathcal{-E}_\textrm{K}/k_\textrm{B}T_\textrm{e}}$, with the electron temperature $T_\textrm{e}$ determined from the cycle-averaged ponderomotive potential. It is to be noted that 2D PIC simulations are known to overestimate the electrons energies by a factor in the range of 1.5-2\cite{Liu2013,Stark2017} due to the electron confinement in a 2D plane, which makes them feel stronger electric fields. This explains why the observed cutoff energies are higher than what is reported experimentally. Nevertheless, it is possible to overcome this effect by comparing ratios of energies, namely using the cutoff energies of the forward electrons for the two intensities $\mathcal{E}_\textrm{max}^\textrm{PIC}(a_0 = 4.86) = 4.9$ MeV and $\mathcal{E}_\textrm{max}^\textrm{PIC}(a_0 = 3.08) = 2.7$ MeV, yielding:

\begin{equation} \label{PICeq}
\Delta\mathcal{E}_\textrm{relative} = 1-\frac{\left[\mathcal{E}_\textrm{max}^\textrm{PIC}(a_0 = 4.86)/\mathcal{E}_\textrm{max}^\textrm{PIC}(a_0 = 3.08)\right]}{\left[\overline{\mathcal{E}}_{\textrm{max}}^\textrm{pond}(a_0 = 4.86)/\overline{\mathcal{E}}_{\textrm{max}}^\textrm{pond}(a_0 = 3.08)\right]} = 3.7\%
\end{equation}

\noindent where $\overline{\mathcal{E}}_{\textrm{max}}^\textrm{pond}(a_0 = 4.86)$ and $\overline{\mathcal{E}}_{\textrm{max}}^\textrm{pond}(a_0 = 3.08)$ are the cycle-averaged ponderomotive energies of 1.32 MeV and 0.71 MeV, respectively. The good agreement with the ponderomotive energy scaling further show that the ponderomotive force is the dominant acceleration mechanism in this configuration. Regarding the backward accelerated electrons, we note that their energies are roughly 2 times lower and that there are about 5 times fewer electrons. The forward-traveling electrons are pushed by the leading edge of the pulse through the ponderomotive force $\boldsymbol{F}_\textrm{pond} \propto \nabla(\vert\boldsymbol{E}\vert^2)$ and the $\boldsymbol{v}\times\boldsymbol{B}$ forward push from the relativistic intensity regime, whereas the backward-traveling feel only the ponderomotive force from the trailing edge of the laser pulse. Hence, it is important to note that it is the relativistic ponderomotive force that enables the formation of a directional, forward-oriented conical beam of electrons, as observed in this experiment. This regime is reached for intensities above $I_0 = 4\times10^{17}$ W/cm$^2$ with $\lambda_0 = 1.8$ $\mu$m.\\

\section{HVL determination}
\noindent

The deposited dose $D$ of a photon beam going through a material is written as:

\begin{equation} \label{dose}
D = \frac{\mu_\textrm{en}}{\rho} \, \mathcal{E} \, \phi 
\end{equation}

\noindent where $\frac{\mu_\textrm{en}}{\rho}$ is the mass energy absorption coefficient of the material, $\mathcal{E}$ is the photon energy and $\phi=\frac{dN_\textrm{ph}}{dA}$ is the photon fluence (\textrm{i.e.}, the number of photons $N_\textrm{ph}$ per unit area $A$). The attenuation of a collimated, monochromatic X-ray photon beam with energy $\mathcal{E}_0$, passing through a uniform material of thickness $x$ and undergoing a negligible amount of scattering (\textit{i.e.} keV-ranged) follows the well-known Beer-Lambert law:

\begin{equation} \label{BL1}
D(x) = D_0 e^{-\mu \, x}
\end{equation}

\noindent with $D_0$ being the incident dose at $x=0$ and $\mu$ is the total attenuation coefficient. The Half-Value Layer (HVL) is the depth at which the dose decreases to half its incident value at $x=0$, therefore when $D(x_{1/2}) = D_0/2$ and yields HVL $= x_{1/2} = \ln(2)/\mu$, with $\mu = \mu(\mathcal{E}_0)$. For a polychromatic photon beam, this is expressed as:

\begin{equation} \label{BL2}
\frac{D(x_{1/2})}{D_0} =\frac{\displaystyle\int \frac{\mu_\textrm{en}}{\rho}(\mathcal{E}) \, \mathcal{E} \, \frac{d\phi}{d\mathcal{E}}(\mathcal{E}) \, e^{-\mu(\mathcal{E}) \, x_{1/2}} \, \D \mathcal{E}}{\displaystyle\int\frac{\mu_\textrm{en}}{\rho}(\mathcal{E}) \, \mathcal{E} \, \frac{d\phi}{d\mathcal{E}}(\mathcal{E}) \, \D \mathcal{E}} = \frac{1}{2}
\end{equation}

\noindent with $\frac{d\phi}{d\mathcal{E}}(\mathcal{E})$ defining the fluence spectrum. Due to the integral formulation of the equation \eqref{BL2}, it is not possible to directly solve for $x_{1/2}$ to find a useful HVL expression. If we evaluate the photon spectrum as a Dirac function centered at an effective photon energy $\mathcal{E}_\textrm{eff}$, \textit{i.e.} $\mathlarger{\frac{d\phi}{d\mathcal{E}}}(\mathcal{E}) = \phi_0 \, \delta(\mathcal{E}-\mathcal{E}_\textrm{eff})$, this yields:

\begin{equation} \label{BL3}
\frac{D(x_{1/2})}{D_0} = e^{-\mu(\mathcal{E}_\textrm{eff}) \, x_{1/2}} = \frac{1}{2} \hspace{1cm} \implies \hspace{1cm} x_{1/2} = \textrm{HVL} = \frac{\ln(2)}{\mu(\mathcal{E}_\textrm{eff})}
\end{equation}

\noindent where $\phi_0$ is the total fluence as:

\begin{equation} \label{phitot}
\phi_0 =  \displaystyle\int \frac{d\phi}{d\mathcal{E}}(\mathcal{E}) \, \D \mathcal{E}
\end{equation}

\noindent The effective photon energy $\mathcal{E}_\textrm{eff}$ corresponds to:

\begin{equation} \label{Eave}
\mathcal{E}_\textrm{eff} = \frac{1}{\frac{\overline{\mu_\textrm{en}}}{\rho} \phi_0} \displaystyle\int \frac{\mu_\textrm{en}}{\rho}(\mathcal{E}) \mathcal{E} \, \frac{d\phi}{d\mathcal{E}}(\mathcal{E}) \, \D \mathcal{E}
\end{equation}

\noindent with $\frac{\overline{\mu_\textrm{en}}}{\rho}$ being the mass energy absorption coefficient averaged over the fluence spectrum as:

\begin{equation} \label{mu0}
 \frac{\overline{\mu_\textrm{en}}}{\rho} =\frac{1}{\phi_0} \displaystyle\int \frac{\mu_\textrm{en}}{\rho}(\mathcal{E}) \frac{d\phi}{d\mathcal{E}}(\mathcal{E}) \, \D \mathcal{E}
\end{equation}

\noindent  Note that the effective energy is not the mean energy of the spectrum, but rather the energy equivalent to a monochromatic beam having the same HVL, which is straightforward to evaluate experimentally.\\


